\documentclass[twocolumn,aps,pra,showpacs,superscriptaddress,
longbibliography,groupedaddress,10pt]{revtex4-1}

\usepackage{bm}
\usepackage{amssymb}
\usepackage[usenames,dvipsnames]{color}
\usepackage{graphicx}
\usepackage{dcolumn}
\usepackage{hyperref}
\usepackage{natbib}
\usepackage[caption=false]{subfig}
\usepackage{float}
\usepackage{times}
\usepackage{booktabs}		%linjer i tabeller...
\usepackage{mathtools}
\usepackage{amsmath,array}
\usepackage{relsize}
\usepackage{enumerate}
\usepackage{color}

\usepackage{listings}
\usepackage{epstopdf}
\usepackage{braket}

\graphicspath{{graphics/}}
\begin{document}

\title{Non-Sequential Double Recombination High Harmonic Generation in Molecular-like Systems}
\author{Kenneth K. Hansen}
\author{Lars Bojer Madsen}
\address{Department of Physics and Astronomy, Aarhus University, DK-8000, Denmark}

%Frequency conversion (nonlinear optics), Multiphoton ionization and excitation molecular spectra, Electron-molecule collisions elastic scattering

\begin{abstract}
We present a study of non-sequential double recombination (NSDR) high harmonic generation (HHG) in a molecular-like system.
Using a Coulomb-corrected three-step model we are able to classically describe the observed NSDR HHG cutoffs precisely for all internuclear distances showing an intrinsic dependence on the location of the nuclei in the NSDR HHG process originating partly from the strong electron correlation in the NSDR HHG process. This dependence modifies the classically allowed return energies which in return changes the cutoffs observed in the HHG spectra.
We also observe that the NSDR HHG signal changes its character for internuclear distances of $R\gtrsim 8-9$ a.u. which is proposed to stem from a change in the charge transfer dynamics within the molecule.
For large internuclear distances of $R\gtrsim 13$ a.u. we observe a clear signature of the point of emission for the first electron emitted in the NSDR HHG signal and we also see signs of molecular exchange paths contributing to the HHG spectrum for these internuclear distances.
\end{abstract}
\date{\today}
\maketitle

\section{Introduction}\label{intro}
Attosecond physics has expanded its versatility greatly over the last decades and the systems studied and the processes of interest have both grown with it \cite{atto_review}. One of the goals of ultrafast physics is the study of molecular and atomic systems on their natural timescales including electron dynamics \cite{atto_review_new}. One such process that has been used for investigating ultrafast dynamics is high harmonic generation (HHG) (see, e.g. Ref. \cite{HHG_measurement_techniques} for a recent example). One-electron HHG has been found to be in general adequately described by the three-step model \cite{Krause_1992,Corkum_HHG,Lewenstein_HHG}. In this model an electron tunnels out from an atom or molecule because of an interaction with a field and propagates in the field until recollision. At recollision a single photon is emitted with energy, $\Omega$, equal to the sum of the ionization potential, $I_p$, and the kinetic energy upon recollision, $K$, i.e., $\Omega = I_p + K$.
HHG has many usages including the creation of attosecond pulses \cite{Isolated_atto_pulse,atto_pulse_creation2}.
A decade ago, a two-electron effect in HHG was identified \cite{NSDR}. In this process two electrons combine their return kinetic energy into a single photon now with a total energy of $\Omega = K_1 + K_2 + I_p^{(1)} + I_p^{(2)}$ where $I_p^{(i)}$ is the $i$'th electrons ionization potential and $K_i$ is the return kinetic energy of the $i$'th electron, thereby reaching even higher photon energies than what is expected in the standard one-electron HHG process \cite{Corkum_HHG,Lewenstein_HHG}.
This non-sequential double recombination (NSDR) process has a distinctive contribution to the HHG signal as a new plateau in the HHG spectrum that emerges at higher photon energies with multiple new cutoffs.
The original work on NSDR HHG \cite{NSDR} found that for the atomic system electrons returning to emit NSDR HHG had to be emitted in different half periods of the external field as electrons emitted in the same half period would have to propagate the same path to return at the same time, a process which is strongly suppressed by electron-electron repulsion. Later work found that for molecules the possibility for emission of electrons from the different nuclei at large internuclear distances reduced the effect of the electron-electron repulsion allowing for same-period emission and recombination (SPEAR) of the electrons creating a new cutoff in the spectrum \cite{SPEAR}. It was also found that the strong correlation of the electrons in the process lead to SPEAR NSDR HHG being dependent on the internuclear distance because of an interaction with nuclei when the electrons propagated in the field.

In this paper we present a study of the NSDR HHG signal and generalize the findings found for SPEAR NSDR to the general case of NSDR HHG in atomic or molecular systems and formulate a Coulomb corrected three-step model (CC-TSM) for two electrons and one or more nuclei. This CC-TSM is able to predict the cutoff energies observed in the NSDR HHG spectrum with unmatched precision.

We find that for small internuclear distances of $R\lesssim 8$ a.u. the observed cutoffs in the NSDR HHG part of the spectra remain similar to the atomic ones as is also predicted by our CC-TSM. For internuclear distances of $R\simeq$ 8-9 a.u. we find a change in the position of the cutoff which matches the predicted shift in the proposed CC-TSM originating from a transition in the charge transfer dynamics in the molecule. After this shift a large discrepancy is observed between the conventional three-step model and the CC-TSM. The three-step model fails to predict the observed cutoffs while the CC-TSM shows the dynamical change in the cutoff for the molecular system as the internuclear distance grows. Finally a clear indication of molecular exchange paths, i.e., paths that start at one nucleus and end at another, is seen in the spectrum for large internuclear distances.

The paper is organized as follows. In Sec. \ref{classical} we present the CC-TSM for NSDR HHG in both atomic and molecular systems. In Sec. \ref{results} we present HHG spectra obtained by time-dependent Schr\"{o}dinger equation  (TDSE) calculations and compare them with the classically calculated cutoffs found in the CC-TSM. Section \ref{sum} summarizes the results of this paper and discusses future points of interest for NSDR HHG.
We use atomic units throughout the paper unless otherwise indicated.

\begin{figure}
\centering
\includegraphics[width=\columnwidth]{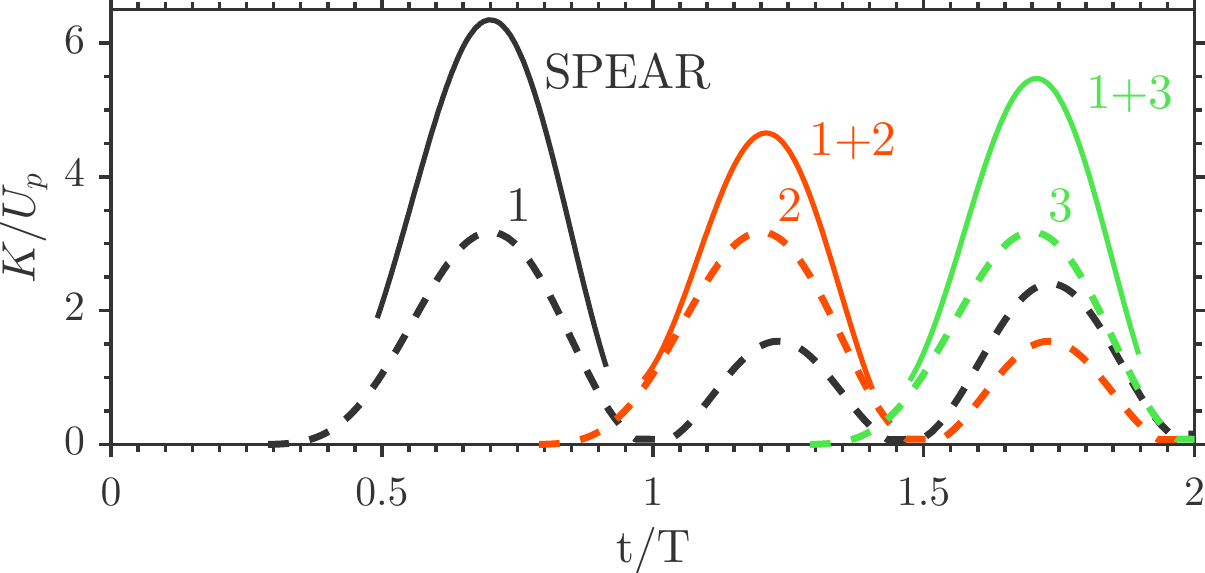}
\caption{Classical return kinetic energies of electrons emitted in the first half period (1, black, dashed), second half period (2, orange, dashed) and third half period (3, green, dashed). The non-sequential double recombination return kinetic energies are shown as full lines above the dashed lines. Same-period emission and recombination energies are indicated by SPEAR; an electron returning for the first time added with an electron returning the second time  is indicated by "1+2" and likewise an electron returning for the first time combined with an electron returning for the third time is indicated by "1+3".}
\label{fig:non-pulse_HHG}
\end{figure}

\section{Coulomb-corrected three-step model}\label{classical}
The standard three-step model assumes that the electron active in the HHG process is emitted from the system at time $t_i$ with velocity $\mathbf{v}(t_i) = 0$ at $\mathbf{r}(t_i)=0$ \cite{Corkum_HHG}. It then propagates in the external field with the electron momentum given as:
\begin{equation}
\mathbf{p}(t) = \mathbf{A}(t) -\mathbf{A}(t_i),
\label{eq:hhg_momentum}
\end{equation}
where $\textbf{A}(t)$ is the vector potential defined by the external field, $\mathbf{F}(t)$, as $\mathbf{F}(t) = - \partial \textbf{A} /\partial t$. The position is then found by integration as:
\begin{equation}
\mathbf{r} (t) = \pmb{\alpha}(t) - \pmb{\alpha} (t_i) - \textbf{A}(t_i)(t-t_i)
\label{eq:hhg_position}
\end{equation}
where $\pmb{\alpha}(t) = \int_{t_i}^t dt' \textbf{A}(t')$ is the quiver motion of the electron in the field. Calculating the return times, $t_r$, of the electron where $\mathbf{r}(t_r)=0$, one can then obtain the return kinetic energy and doing this for all emission times the highest return kinetic energy can be obtained for a given pulse.
Using a continuous wave one finds the well known maximum return kinetic energy of $3.17 U_p$ where $U_p$ is the ponderomotive energy defined as $U_p = F_{0}^2/ 4 \omega^2$, with $\omega$ being the angular frequency of the driving field with maximum amplitude $F_0$. For NSDR HHG it was previously found that electrons returning for a second or third time give an important contribution to the NSDR HHG signal \cite{NSDR}. This can be seen in Fig. \ref{fig:non-pulse_HHG} where the return kinetic energy is shown as a function of the return time for multiple periods of emission.
It is noted that the second and third return of electrons cannot reach the same return kinetic energies as the first return and because of the small window of time in the pulse where electrons returning several times are emitted their contribution to the signal is normally not observable in an HHG spectrum.
To obtain the cutoffs for NSDR HHG in this model we add up each electrons return kinetic energy giving us predictions of the cutoff energies observable in the NSDR HHG part of the spectra \cite{NSDR}. This is indicated in Fig. \ref{fig:non-pulse_HHG} by the "1+2" for an electron returning for the first time and another electron returning for the second time and "1+3" for an electron returning the first time and another electron returning for the third time.
SPEAR NSDR is also indicated in Fig. \ref{fig:NSDR_cutoff} which was previously shown to be relevant for large internuclear distances, here calculated by adding up the first return of an electron twice.
For more discussion about these cutoffs without Coulomb correction see Ref. \cite{NSDR,SPEAR}.
In Ref. \cite{SPEAR} we found that an interaction with the bare nuclei was necessary to correctly predict the behavior of SPEAR NSDR HHG and we will therefore now introduce this effect of nuclei interacting with electrons when the electrons are propagating past bare nuclei on their first and second pass over the nuclei in "1+2" and "1+3" NSDR HHG. This interaction will introduce a characteristic dependence on the internuclear distance in the NSDR HHG signal and it will later be seen that this dependence is indeed present in the actual HHG signal.
\begin{figure}
\centering
\includegraphics[width=0.7\columnwidth]{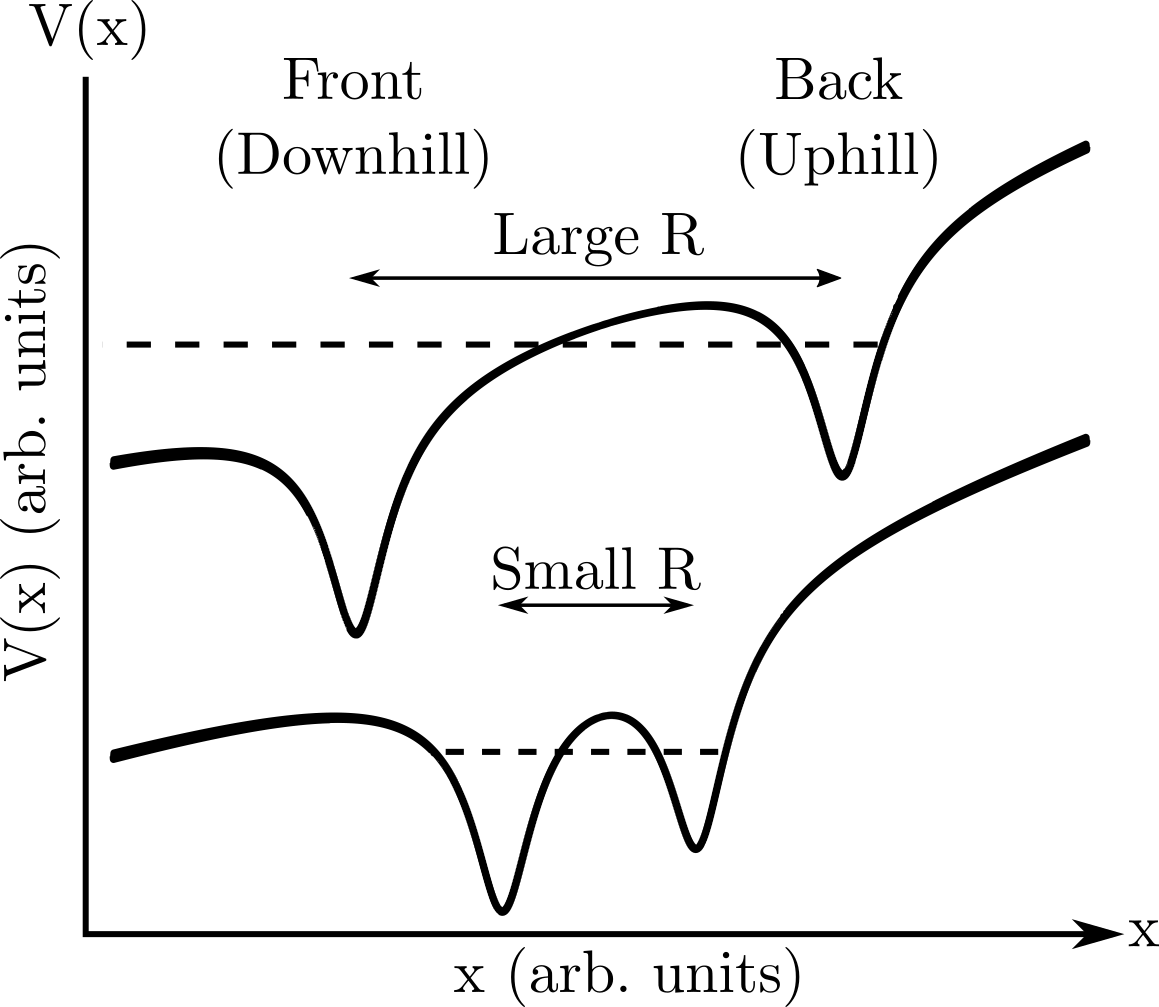}
\caption{Illustration of the Coulomb potentials for large and small internuclear distances ($R$) for a particular value of the electric field pointing to the right. The left minimum is the front (downhill) nucleus for emission and the minimum to the right is the back (uphill) nucleus for emission.
The dashed lines signify how an electron emitted at the back nucleus in a molecule with a large internuclear distance can propagate away freely  after tunneling without being caught in the front nucleus' minimum while for small internuclear distance the electron cannot escape from the front nucleus' minimum classically. The potential of the front nucleus therefore should not be included in the first part of the classical propagation for small internuclear distances. See text.}
\label{fig:front_back_ill}
\end{figure}
\begin{figure}
\centering
\includegraphics[width=0.8\columnwidth]{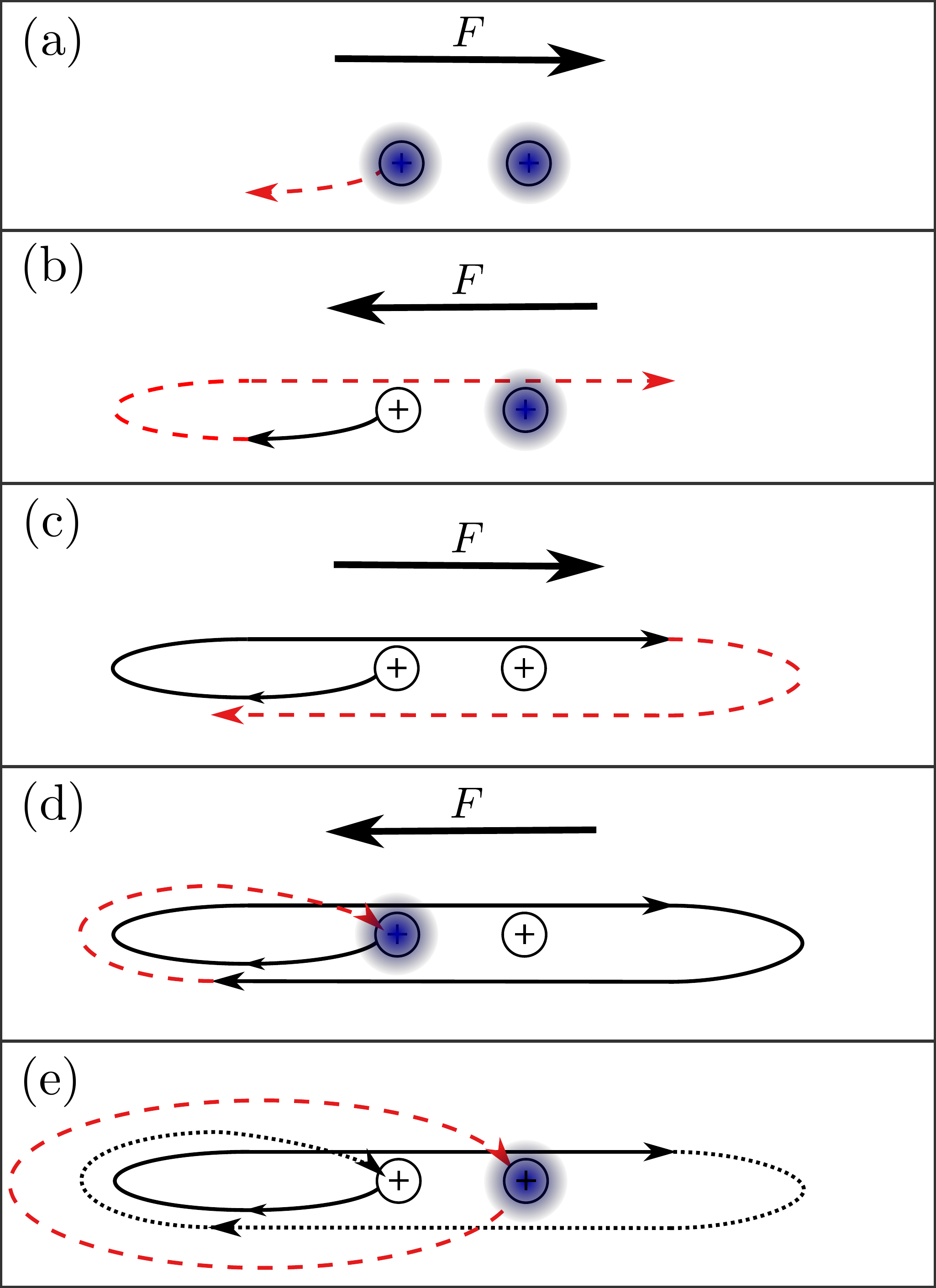}
\caption{Illustration of the path and interactions for electrons in "1+3" NSDR in a molecule with large internuclear distance. 
(a)-(d) the path for an electron emitted first in the NSDR process. The field direction changes in each panel and with it the effective Coulomb interaction felt by the electrons changes. The path of the electron with the Coulomb interaction and electric field shown in the panel is shown as (red) dashed. The path of the electron for previous periods is shown as solid black.
In the Coulomb corrected three-step model the Coulomb interaction from each nucleus in H$_2$ is included in the propagation step but only at certain times. The (blue) cloud covering a nucleus signifies that this nucleus does not interact with the electron on the (red) dashed part of its path.
(e) a path of the electron emitted as the second one in the NSDR process so therefore the illustration is for a molecule with a large internuclear distance. Therefore the electron interacts with a single nucleus throughout its entire path and the Coulomb interactions do not change when the field changes direction. The dashed (red) curve is the path of the electron emitted second in the NSDR HHG process. The (black) solid curve is the path of the first electron before the second electron is emitted and the (black) dotted line is the path of the first electron after the second electron is emitted.
The field changes direction during the time evolution in (e) and is therefore not shown.}
\label{fig:TD_coulomb_interaction_example}
\end{figure}

The following method will be independent on pulse frequency and amplitude but the results will be dependent on these variables. We consider a homonuclear model of the H$_2$ molecule where the molecular axis is co-linear with the polarization of the field. To isolate the physics related to the considered CC-TSM and not to be concerned with carrier-envelope pulse effects, we use a continuous wave of the type $F(t) = F_0 \cos(\omega t)$ where $F_0 = 0.119$ $(5\times 10^{14} \text{W/cm}^2)$ and $\omega = 0.0584$ $(\lambda = 780 $nm).
We still use the first (ionization) and last (recombination) steps of the three-step model in our CC-TSM but change the propagation step to include interactions with the bare nuclei as described below.
We still assume that the electron is emitted at a nuclei at $r(t_i) = 0$ with velocity $v_i = 0$ but we modify the equations of motion to include an interactions with the nuclei. 
The TDSE calculations presented later are performed in a reduced dimensionality model of H$_2$ and a softened Coulomb is therefore used. We use the same interaction in the quantum mechanical and in the classical treatments. The softened Coulomb potential has the form:
\begin{align}
V(x) = -\frac{1}{\sqrt{(x - R)^2 + \epsilon_{ei}}} -\frac{1}{\sqrt{x^2 + \epsilon_{ei}}},
\label{eq:coulomb_potential}
\end{align}
where $x$ is the position of the electron, $\epsilon_{ei} = 0.329$ is the softening used and the left and right term is included in the description of the propagation of the electron depending on the instantaneous position of the electron. In this way the Coulomb interaction with the nuclei will be dependent on what type of NSDR that is modeled: the Coulomb interaction will be different for the first electron emitted in "1+2" NSDR compared with the first electron emitted in "1+3" NSDR.
There will also be a dependence in the return kinetic energy from electrons with different points of emission. In H$_2$ the electrons can be emitted from the different nuclei and these two points will have differing interactions for the electron. Depending on the orientation of the field, the electron either has to pass over a nucleus after emission or just propagate out freely
therefore creating a front and a back nucleus of emission as defined in Fig. \ref{fig:front_back_ill}. For "1+2" and "1+3" NSDR there will be two different return energies depending on whether the first electron is emitted from the front or back nucleus compared to the orientation of the field.

The Coulomb interaction with an electron also depends on the internuclear distance. For small distances if an electron is emitted with zero velocity it will classically continue to be caught in the Coulomb potential from the non-emission nucleus as is illustrated by the dashed line in the potential curve for small $R$ seen in Fig. \ref{fig:front_back_ill}.
We will therefore start the Coulomb interaction at a later point in time and space when the electron is free of the molecule and able to propagate away.
For large internuclear distances the electron is not classically caught as illustrated by the dashed line in the potential curve for large $R$ seen in Fig. \ref{fig:front_back_ill}. It was also shown earlier that the Coulomb interaction for the electron emitted from the back nuclei was important for large internuclear distances in the SPEAR NSDR HHG process\cite{SPEAR}.

\begin{figure}
\centering
\includegraphics[width=\columnwidth]{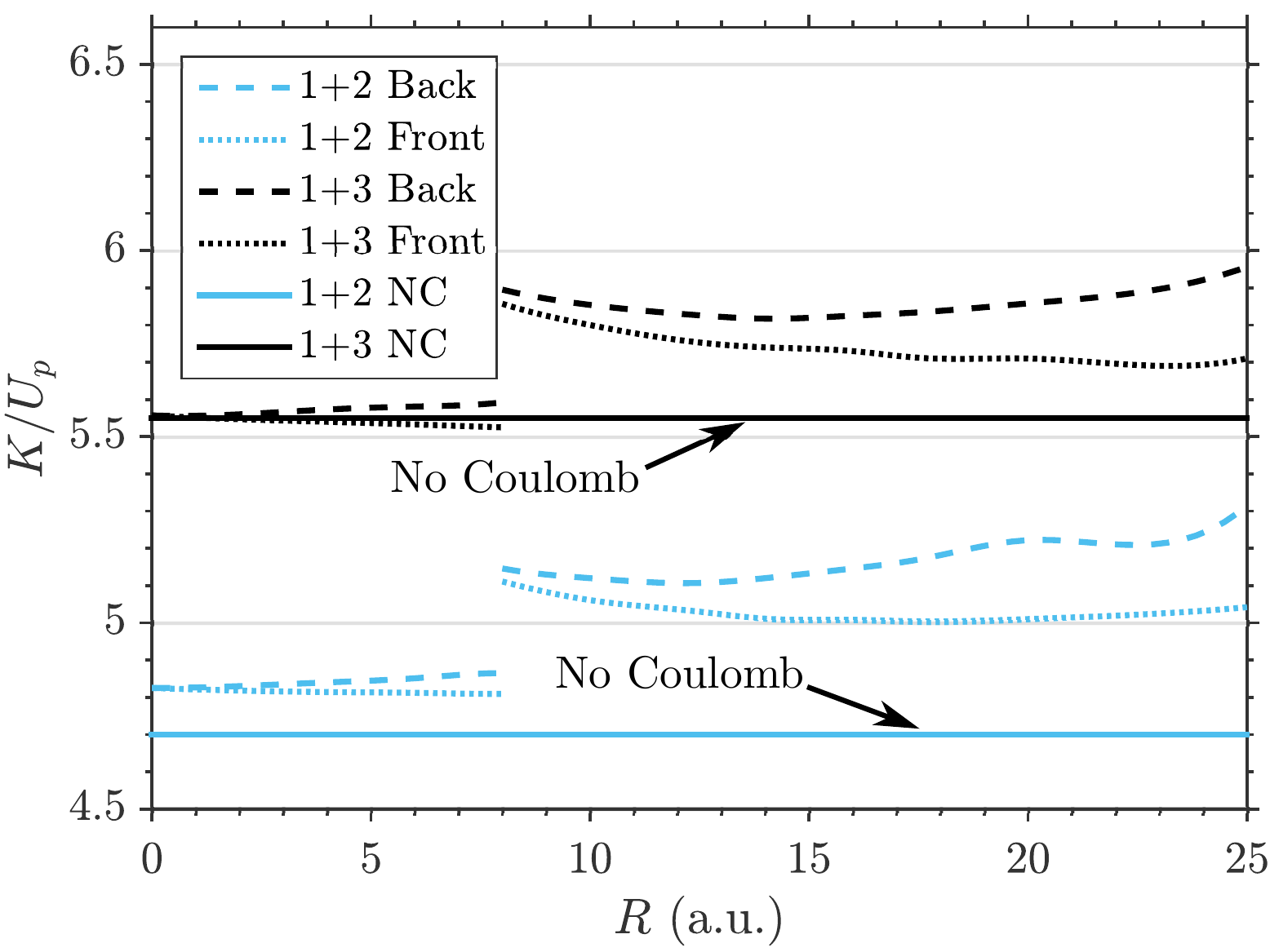}
\caption{Maximum classical return kinetic energies for "1+2" and "1+3" NSDR for front or back emission as a function of the internuclear distance. The normal three-step model cutoff energies are shown as references indicated by "No Coulomb" (NC). See Fig. \ref{fig:front_back_ill} for definition of front and back.}
\label{fig:NSDR_cutoff}
\end{figure}

As an example of how the Coulomb interaction changes with the electron trajectory, and therefore with time, we present in detail the time-dependent Coulomb interaction for the case of "1+3" NSDR where the first electron is emitted from the front nucleus in a molecule with a large internuclear distance.
In Fig. \ref{fig:TD_coulomb_interaction_example} (a) the electron has just been emitted from the front nucleus in the molecule and does not interact with any of the nuclei, represented by the (blue) clouds covering each nuclei.
In Fig. \ref{fig:TD_coulomb_interaction_example} (b) the field changes direction and the electron propagates back across the molecule interacting with one bare nucleus as the second electron has not been emitted at this point in the "1+3" NSDR process.
In Fig. \ref{fig:TD_coulomb_interaction_example} (c) the field changes direction again, and the electron propagates across the molecule once more now interacting with both bare nuclei as the second electron has been emitted at this point in the "1+3" NSDR process.
Finally in Fig. \ref{fig:TD_coulomb_interaction_example} (d) the electron returns to the nucleus from which it was emitted interacting only with the non-emission nucleus as interacting with the emission nucleus would be adding the ionization potential twice.
In Fig. \ref{fig:TD_coulomb_interaction_example} (e) the interaction for the electron emitted second is shown for the entire propagation. Because we assume a large internuclear distance the electron interacts with the non-emission nucleus during its entire propagation but never interacts with the nucleus from which it was emitted.
Similar steps have been used to calculate classical return kinetic energies for the different types of NSDR HHG. The calculated return kinetic energies for "1+2" and "1+3" NSDR in the cases where the first electron is emitted from either the front or back nucleus can be seen in Fig. \ref{fig:NSDR_cutoff}. The results of the standard three-step model are indicated by the "No Coulomb" (NC) lines in the figure which differ from the CC-TSM already at $R=0$ in the case of "1+2" NSDR.
There is a discontinuity in Fig. \ref{fig:NSDR_cutoff} at $R=8$ which is the point where the system changes from small internuclear distances where the Coulomb interaction from the non-emission nucleus is excluded at the start of the propagation to large internuclear where the interaction with the non-emission nucleus is included as was seen in Fig. \ref{fig:TD_coulomb_interaction_example} (e).
This distance of $R=8$ is determined from the TDSE calculations presented later.
It is also observed that the cutoff for the "1+2" front and back NSDR become energetically different for large internuclear distances creating a distinction for their contribution to the spectrum and likewise for "1+3" NSDR.

The cutoff return kinetic energies obtained with the CC-TSM will now be compared with TDSE calculations. This comparison serves to validate our classical two-electron, two-nuclei CC-TSM.

\section{Numerical results}\label{results}
We have solved the TDSE for a reduced dimensionality model of H$_2$. This was done using a split-step operator Crank-Nicolson (Peaceman-Rachford) method as in Ref. \cite{NSDR}. The Coulomb potential used is given in Eq. (\ref{eq:coulomb_potential}) and the softening factors for the Coulomb potentials were chosen such that the ionization potential for He$^{+}$ was $I_p (\text{He}^+) = 2.0$ and $I_p (\text{He}) = 0.903$ for He. For more details about the model and numerical method see Ref. \cite{SPEAR}.
The pulse used was a 21-cycle pulse with 15-cycles of the form $A(t) = F_0/\omega \sin(\omega t)$ with a 3-cycle sinusoidal ramp-up and ramp-down, where $F_0 = 0.119$ and $\omega = 0.0584$. The same field parameters were used in the classical analysis.
The dipole acceleration was calculated in each time-step and the harmonic spectrum was found from the norm square of the Fourier transform of the dipole acceleration i.e. $|\tilde{a}_{dip} (\Omega)|^ 2$.
We have performed calculations for a large range of internuclear distances ($R\in [0,80]$) and will now show representative results illustrating the trends and conclusions that can be drawn from the data and analysis.

\begin{figure}
\centering
\includegraphics[width=\columnwidth]{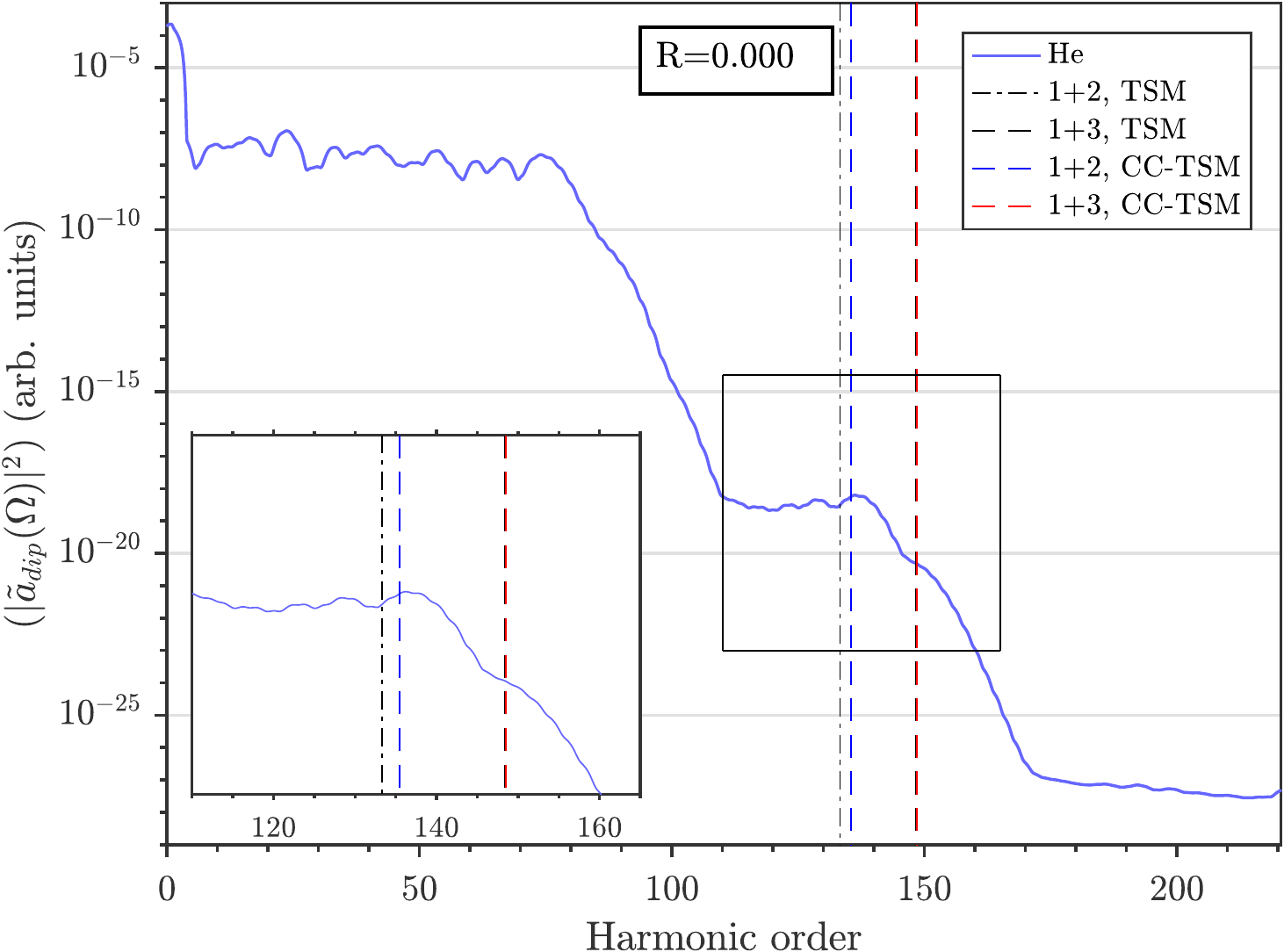}
\caption{The dipole acceleration spectrum for He interacting with a 21-cycle laser pulse at $\omega = 0.0584$ $(\lambda = 780$nm) and $F_0 = 0.119$ $(5\times 10^{14} \text{W/cm}^2)$. The insert to the lower-left is a zoom-in of the box around the second plateau of the spectrum. The vertical lines indicate the classical cutoff energies for the three-step model (TSM) and the Coulomb corrected three-step model (CC-TSM) for "1+2" and "1+3" NSDR HHG (see text).}
\label{fig:Helium}
\end{figure}

In Fig. \ref{fig:Helium} the dipole acceleration spectrum for He is shown. The spectrum has been smoothed to highlight the underlying structures as the position of the cutoff is of main interest for the comparison with the classical models.
In the spectrum there are two plateaus where the low energy one is the one-electron HHG signal with the normal cutoff at $\Omega = 3.17U_p + I_p$.
In this context we define a cutoff as a point in the spectrum where the spectrum reaches a peak and then falls off exponentially over several amplitudes. Such a cutoff physically originates from classical electron paths only being able to reach the maximum return kinetic energy at a certain time in the pulse therefore creating a peak for then to fall off exponentially as there are no more classically allowed paths at higher energies. Such an effect will be seen for all types of HHG if the signal stands clear from other contributions to the signal.
The second plateau at higher energy, but of lower magnitude, is the NSDR HHG plateau which can be observed to have two cutoffs. The insert in Fig. \ref{fig:Helium} is a zoom of the box around the NSDR plateau in the spectrum wherein the positions of the vertical lines can be seen more clearly. 
From the left the first dashed (blue) and second dashed (red) lines are positioned at the maximum return kinetic energy for both electrons in the "1+2" CC-TSM and "1+3" CC-TSM models respectively. The (black) dashed-dotted line is the "1+2" cutoff in the conventional three-step model (Eqs. \ref{eq:hhg_momentum}-\ref{eq:hhg_position}) located to the left of the "1+2" CC-TSM line while the "1+3" three-step model (black) dashed line is located beneath the (red) dashed "1+3" CC-TSM line. These findings are in accordance with the results in Fig. \ref{fig:NSDR_cutoff}.
In Fig. \ref{fig:Helium} we observe a good agreement between the CC-TSM model and the observed cutoffs. There is no difference for the "1+3" NSDR cutoff between the three-step model and CC-TSM but for "1+2" NSDR the CC-TSM precisely predicts the observed cutoff, clearly setting it apart from the results obtained with the three-step model and underlining the necessity of including the Coulomb interaction in the modeling of NSDR. It is noted that even though the signal from "1+3" NSDR is present in the spectrum a clear cutoff is not seen as the low magnitude of the signal and the proximity to the "1+2" NSDR cutoff only leaves a shoulder in the drop off from the "1+2" NSDR.

\begin{figure}
\centering
\includegraphics[width=\columnwidth]{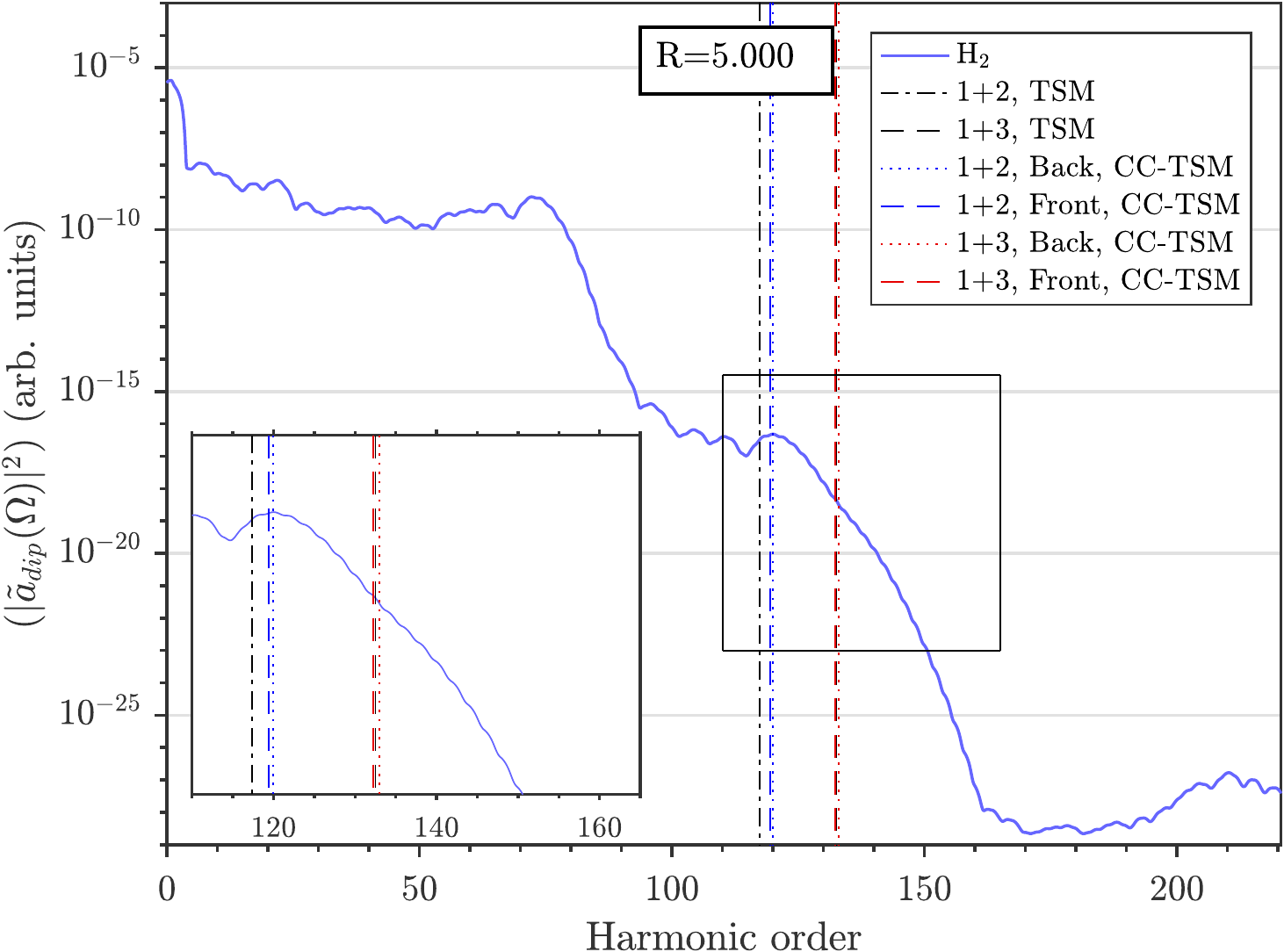}
\caption{The dipole acceleration spectrum for H$_2$ with $R=5.0$ and pulse parameters as in Fig. \ref{fig:Helium}. The vertical lines indicate the classical cutoff energies for the three-step model (TSM) and the Coulomb corrected three-step model (CC-TSM) in the cases of the first electron being emitted from the front or back nuclei in the molecule compared to the orientation of the field in "1+2" and "1+3" NSDR HHG (see text). See caption in Fig. \ref{fig:Helium} for more.}
\label{fig:R5}
\end{figure}

The dipole acceleration spectrum for H$_2$ with an internuclear distance of $R=5.0$ is shown in Fig. \ref{fig:R5}. A general trend for increasing internuclear distance from the case of $R=0.0$ (He) to around $R=7.0$ is an increase in the NSDR HHG signal originating from an increase in the ionization probability with increasing internuclear distance and charge-resonance-enhanced ionization for intermediate internuclear distances of $R\in[1,5]$ \cite{CREI}. Comparing the spectra for He and H$_2$ in Figs. \ref{fig:Helium} and \ref{fig:R5} the same conclusions can be made in the latter case for the cutoffs predicted by the three-step model and CC-TSM. For $R=5.0$ there is a slight difference in cutoff energies in the CC-TSM model depending on whether the first electron is emitted as a front or back electron.
Looking at the insert in Fig. \ref{fig:R5} we observe that the CC-TSM prediction still fits perfectly with the observed cutoff of "1+2" NSDR and the normal three-step model still deviates from the observed cutoff. Likewise we still do not observe a difference between the CC-TSM and the three-step model for "1+3" NSDR. They are still both located directly at the cutoff observed for "1+3" NSDR.

\begin{figure}
\centering
\includegraphics[width=\columnwidth, trim={0cm 1cm 0cm 0cm}]{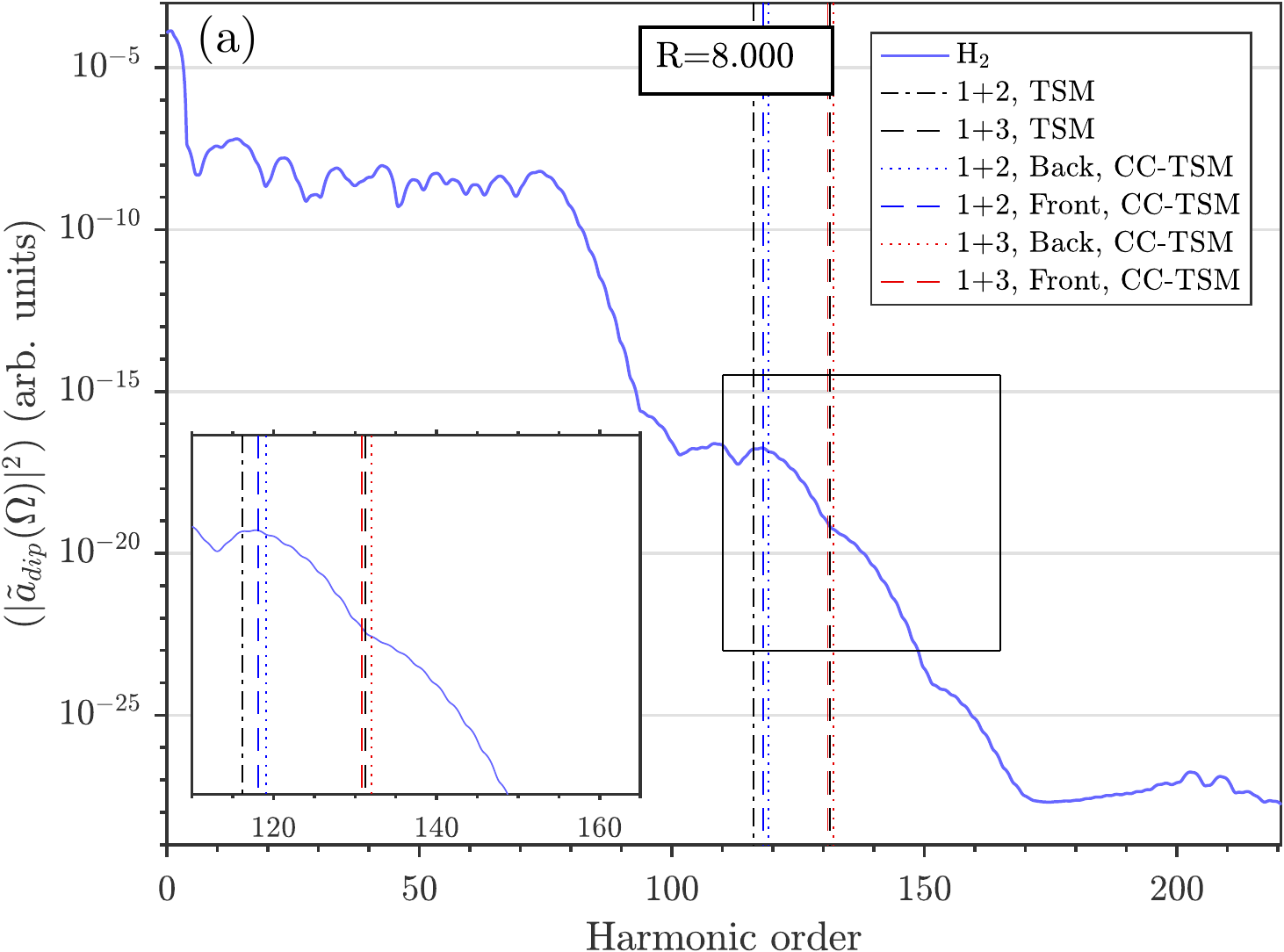}
\includegraphics[width=\columnwidth, trim={0cm 0cm 0cm 0.35cm}]{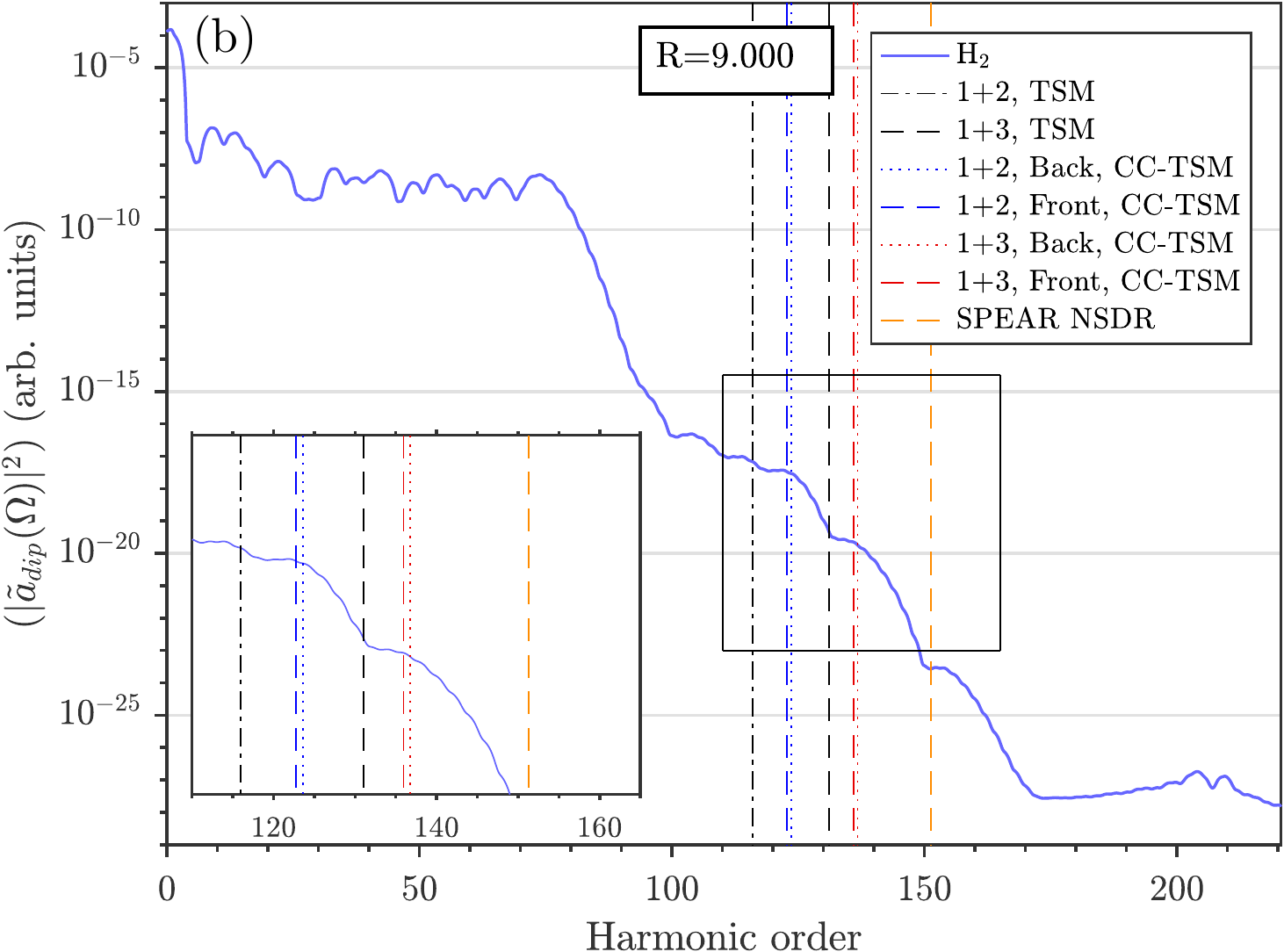}
\caption{The dipole acceleration spectrum for H$_2$ with pulse parameters as in Fig. \ref{fig:Helium} with (a) $R=8.0$ and (b) $R=9.0$. The vertical lines indicate the classical cutoff energies for the three-step model (TSM) and the Coulomb corrected three-step model (CC-TSM).
The dashed (orange) line furthest to the right in (b) is the CC-TSM prediction of the same-period emission and recombination non-sequential double recombination HHG cutoff.
(See caption of Fig. \ref{fig:R5} for definition of the other of vertical lines.)}
\label{fig:R8-9}
\end{figure}

In Figs. \ref{fig:R8-9} (a) and (b) the dipole accelerations for H$_2$ with $R=8.0$ and $R=9.0$ are shown. For these internuclear distances a major shift in the cutoffs are observed as was hinted to in Fig. \ref{fig:NSDR_cutoff}, where a discontinuity is located at these internuclear distances in the predicted cutoffs. In Fig. \ref{fig:R8-9} (a) the vertical lines are the predicted cutoffs from the CC-TSM for an atomic-like system with no interaction with the nuclei in the first half-cycle after emission as was discussed in Sec. \ref{classical}. It was previously found that such an interaction would be necessary to model SPEAR NSDR \cite{SPEAR} and it was included in the classical model used to calculate the cutoff shown in Fig. \ref{fig:R8-9} (b).
Studying the behavior of the HHG spectra in this region of $R=8.0$ to $R=9.0$  a change in the electron dynamics  was found in the molecule around this internuclear distance.
Looking at the NSDR HHG process one electron will be emitted from one nucleus leaving the other electron on the other nucleus. Using wavepacket propagation we find the ground state of an electron on a single nucleus and propagate this with both potentials present.
The electron will then start oscillating between the two nuclei for small internuclear distances, first classically allowed for small distances and then through tunneling for larger internuclear distances. From these charge transfer simulations we observe that around $R\simeq 8-9$ the tunneling time for the electron becomes larger than the oscillation time of the electric field. From these $R$-values and larger, the electron will therefore not have time to delocalize on the two nuclei. This oscillation time for the electron in the molecular system therefore explains the change in the electron dynamic in the classical model, going from a system where the electron effectively sees the system as a whole to a system with separate nuclei for large internuclear distances.
This change in the charge transfer dynamics is clearly seen in the spectra. Looking at Fig. \ref{fig:R8-9} (a) the spectrum is very similar to the ones in Figs. \ref{fig:Helium} and \ref{fig:R5} but for $R=9$ in Fig. \ref{fig:R8-9} (b) the cutoffs predicted by CC-TSM for "1+2" and "1+3" NSDR are now both very different than what is predicted by the normal three-step model, while the CC-TSM predicted cutoffs clearly match the observed cutoffs.
SPEAR NSDR HHG also becomes observable for these internuclear distances but a similar explanation for its occurrence as for "1+2" and "1+3" NSDR HHG is not likely as both electrons are emitted in the same half period in SPEAR NSDR HHG.

\begin{figure}
\centering
\includegraphics[width=\columnwidth]{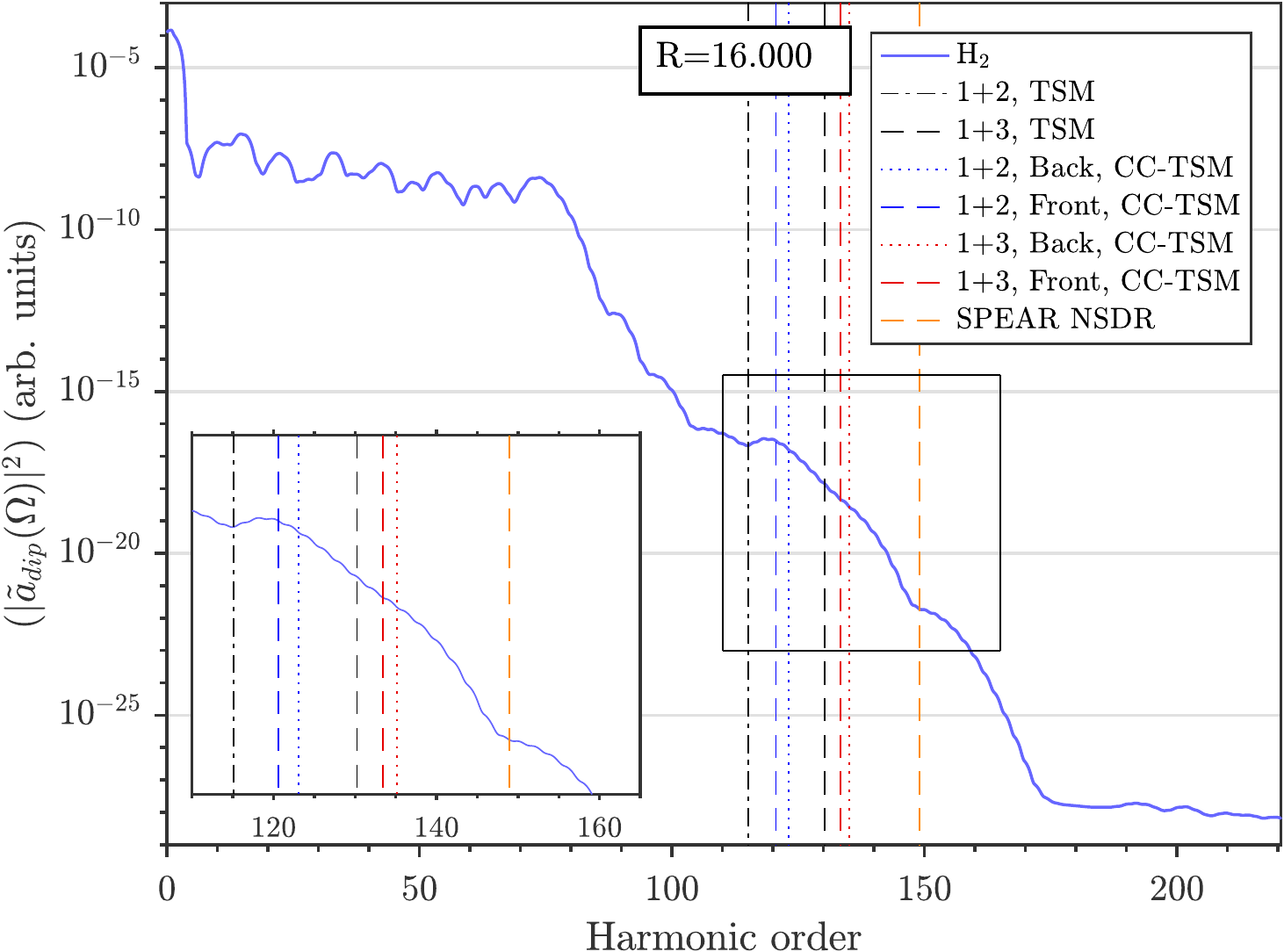}
\caption{The dipole acceleration spectrum for H$_2$ with $R=16.0$ and pulse parameters as in Fig. \ref{fig:Helium}. The vertical lines indicate the classical cutoff energies for the three-step model (TSM) and the Coulomb corrected three-step model (CC-TSM). (See captions of Figs. \ref{fig:R5} and \ref{fig:R8-9} for definition of vertical lines.)}
\label{fig:R16}
\end{figure}

Going to even larger internuclear distance the CC-TSM continues to precisely predict the observed cutoffs but looking at large internuclear distances like in Fig. \ref{fig:R16} (a) for $R=16$ we observe that the clear cutoffs observed for smaller internuclear distance disappear and are replaced by a straight decreasing diagonal line on the logarithmic scale from around 120$\omega$ to 140$\omega$. This effect can be explained by considering molecular exchange paths in the HHG process. When using either the CC-TSM or the normal three-step model we assume that the electron is emitted and recombines at the same nucleus but instead it is possible for the electron to be emitted at one nucleus and recombine at another which results in a different return kinetic energy \cite{Bandrauk}. This change in return kinetic energy will be dependent on the internuclear distance of the specific system and it will be more pronounced for large internuclear distance is the case for $R=16$. These extra paths give rise to extra signals in the energy range with their own cutoff but because of the close proximity in the return energies, they result in a flat linear drop-off instead of the cutoff observed for the smaller internuclear distances.
Observing the "1+2" front CC-TSM shown in Fig. \ref{fig:R16} we observe that this is the predicted cutoff energy that matches the observed cutoff in the spectrum, we can therefore conclude that the first electron will most likely be emitted from the front nucleus in the molecule. This becomes even more apparent for larger internuclear distances where the energy difference between "1+2" NSDR front and back becomes larger and the "1+2" front continues to be located at the cutoff in the spectrum.
The conclusions for $R\simeq16$ continue to be present until $R\gtrsim 25$ where the one-electron signal extends its cutoff energy beyond the NSDR HHG signal due to direct exchange paths.

\section{Summary and Outlook}\label{sum}
In this paper we have performed a study of the NSDR HHG signal from a molecular-like system and presented a special version of the three-step model for the two-electron, two-nuclei case which we called the Coulomb corrected three-step model (CC-TSM). This model is specifically constructed for two-electron NSDR HHG as the strong electron-electron correlation in the NSDR HHG process makes the inclusion of the interaction with the bare nuclei important. We have then used the classical maximum return kinetic energies to predict the cutoffs observed in the HHG spectra showing that the CC-TSM is superior to the ordinary three-step model for all internuclear distances and that the CC-TSM predicts the cutoff precisely for all cases in NSDR HHG.
We observe a change in the NSDR process around internuclear distances of $R\simeq8$ where the observed cutoffs shift their positions abruptly. This shift was proposed to stem from a change in the charge transfer dynamics within the molecule resulting in each nucleus being its own separate entity after emission of the first electron in the NSDR process.
Finally for large internuclear distances of $R\in[13,25]$ we observed that from the position of the cutoff of "1+2" NSDR, we could predict the nucleus from which the electron was most likely to be emitted to be the front nucleus compared with the orientation of the electric field. In this sense the NSDR HHG signal gives very direct access to the initial ionization step in a molecule.
We also observed for these large internuclear distances that exchange paths in the molecule create a continuous drop instead of a cutoff in the NSDR part of the spectrum because of the many contributions to the spectrum in that region.
In this sense our CC-TSM for NSDR HHG has added to the understanding of the NSDR HHG process. For processes where emission and recombination times are crucial such as attosecond pulse-probe experiments this model will be necessary. Attosecond pulses could also be used to increase the yield of the NSDR HHG process opening up for many interesting opportunities for using the strongly correlated two-electron NSDR HHG process to study electron correlation and make precise measurements of molecular dynamics using the NSDR HHG process.

The dependence on the internuclear distance also opens up for opportunities to study molecular dynamics in systems with relatively large internuclear distances where effects such as the previously observed two-center interference in HHG disappear\cite{mol_interferens,HHG_mol_probe1,HHG_mol_probe2}. This two-center interference, normally observed in the one-electron part of the HHG spectrum, can be observed not to be present in the spectrum already at $R\simeq5$ (Fig. \ref{fig:R5}).
The high photon energies in the NSDR HHG signal also makes it stand out separately from the one-electron HHG signal possibly making it easy to isolate experimentally.
We recognize the low magnitude of the NSDR HHG signal but in the future the benefits of the NSDR HHG signal could outweigh the difficulties and make it of experimental interest for studies of molecular systems and electron-electron correlation.

\section*{Acknowledgements}
This work was supported by the VKR Center of Excellence.
We thank Dr. D. Bauer for making the two-electron TDSE program with which much of the calculations were performed available.
The numerical results presented in this work were obtained at the Centre for Scientific Computing, Aarhus.


\begin{thebibliography}{15}%
\makeatletter
\providecommand \@ifxundefined [1]{%
 \@ifx{#1\undefined}
}%
\providecommand \@ifnum [1]{%
 \ifnum #1\expandafter \@firstoftwo
 \else \expandafter \@secondoftwo
 \fi
}%
\providecommand \@ifx [1]{%
 \ifx #1\expandafter \@firstoftwo
 \else \expandafter \@secondoftwo
 \fi
}%
\providecommand \natexlab [1]{#1}%
\providecommand \enquote  [1]{``#1''}%
\providecommand \bibnamefont  [1]{#1}%
\providecommand \bibfnamefont [1]{#1}%
\providecommand \citenamefont [1]{#1}%
\providecommand \href@noop [0]{\@secondoftwo}%
\providecommand \href [0]{\begingroup \@sanitize@url \@href}%
\providecommand \@href[1]{\@@startlink{#1}\@@href}%
\providecommand \@@href[1]{\endgroup#1\@@endlink}%
\providecommand \@sanitize@url [0]{\catcode `\\12\catcode `\$12\catcode
  `\&12\catcode `\#12\catcode `\^12\catcode `\_12\catcode `\%12\relax}%
\providecommand \@@startlink[1]{}%
\providecommand \@@endlink[0]{}%
\providecommand \url  [0]{\begingroup\@sanitize@url \@url }%
\providecommand \@url [1]{\endgroup\@href {#1}{\urlprefix }}%
\providecommand \urlprefix  [0]{URL }%
\providecommand \Eprint [0]{\href }%
\providecommand \doibase [0]{http://dx.doi.org/}%
\providecommand \selectlanguage [0]{\@gobble}%
\providecommand \bibinfo  [0]{\@secondoftwo}%
\providecommand \bibfield  [0]{\@secondoftwo}%
\providecommand \translation [1]{[#1]}%
\providecommand \BibitemOpen [0]{}%
\providecommand \bibitemStop [0]{}%
\providecommand \bibitemNoStop [0]{.\EOS\space}%
\providecommand \EOS [0]{\spacefactor3000\relax}%
\providecommand \BibitemShut  [1]{\csname bibitem#1\endcsname}%
\let\auto@bib@innerbib\@empty
%</preamble>
\bibitem [{\citenamefont {Krausz}\ and\ \citenamefont
  {Ivanov}(2009)}]{atto_review}%
  \BibitemOpen
  \bibfield  {author} {\bibinfo {author} {\bibfnamefont {Ferenc}\ \bibnamefont
  {Krausz}}\ and\ \bibinfo {author} {\bibfnamefont {Misha}\ \bibnamefont
  {Ivanov}},\ }\bibfield  {title} {\enquote {\bibinfo {title} {Attosecond
  physics},}\ }\href {\doibase 10.1103/RevModPhys.81.163} {\bibfield  {journal}
  {\bibinfo  {journal} {Rev. Mod. Phys.}\ }\textbf {\bibinfo {volume} {81}},\
  \bibinfo {pages} {163--234} (\bibinfo {year} {2009})}\BibitemShut {NoStop}%
\bibitem [{\citenamefont {Calegari}\ \emph {et~al.}(2016)\citenamefont
  {Calegari}, \citenamefont {Sansone}, \citenamefont {Stagira}, \citenamefont
  {Vozzi},\ and\ \citenamefont {Nisoli}}]{atto_review_new}%
  \BibitemOpen
  \bibfield  {author} {\bibinfo {author} {\bibfnamefont {Francesca}\
  \bibnamefont {Calegari}}, \bibinfo {author} {\bibfnamefont {Giuseppe}\
  \bibnamefont {Sansone}}, \bibinfo {author} {\bibfnamefont {Salvatore}\
  \bibnamefont {Stagira}}, \bibinfo {author} {\bibfnamefont {Caterina}\
  \bibnamefont {Vozzi}}, \ and\ \bibinfo {author} {\bibfnamefont {Mauro}\
  \bibnamefont {Nisoli}},\ }\bibfield  {title} {\enquote {\bibinfo {title}
  {Advances in attosecond science},}\ }\href
  {http://stacks.iop.org/0953-4075/49/i=6/a=062001} {\bibfield  {journal}
  {\bibinfo  {journal} {Journal of Physics B: Atomic, Molecular and Optical
  Physics}\ }\textbf {\bibinfo {volume} {49}},\ \bibinfo {pages} {062001}
  (\bibinfo {year} {2016})}\BibitemShut {NoStop}%
\bibitem [{\citenamefont {Kraus}\ \emph {et~al.}(2015)\citenamefont {Kraus},
  \citenamefont {Mignolet}, \citenamefont {Baykusheva}, \citenamefont
  {Rupenyan}, \citenamefont {Horn{\'y}}, \citenamefont {Penka}, \citenamefont
  {Grassi}, \citenamefont {Tolstikhin}, \citenamefont {Schneider},
  \citenamefont {Jensen}, \citenamefont {Madsen}, \citenamefont {Bandrauk},
  \citenamefont {Remacle},\ and\ \citenamefont
  {W{\"o}rner}}]{HHG_measurement_techniques}%
  \BibitemOpen
  \bibfield  {author} {\bibinfo {author} {\bibfnamefont {P.~M.}\ \bibnamefont
  {Kraus}}, \bibinfo {author} {\bibfnamefont {B.}~\bibnamefont {Mignolet}},
  \bibinfo {author} {\bibfnamefont {D.}~\bibnamefont {Baykusheva}}, \bibinfo
  {author} {\bibfnamefont {A.}~\bibnamefont {Rupenyan}}, \bibinfo {author}
  {\bibfnamefont {L.}~\bibnamefont {Horn{\'y}}}, \bibinfo {author}
  {\bibfnamefont {E.~F.}\ \bibnamefont {Penka}}, \bibinfo {author}
  {\bibfnamefont {G.}~\bibnamefont {Grassi}}, \bibinfo {author} {\bibfnamefont
  {O.~I.}\ \bibnamefont {Tolstikhin}}, \bibinfo {author} {\bibfnamefont
  {J.}~\bibnamefont {Schneider}}, \bibinfo {author} {\bibfnamefont
  {F.}~\bibnamefont {Jensen}}, \bibinfo {author} {\bibfnamefont {L.~B.}\
  \bibnamefont {Madsen}}, \bibinfo {author} {\bibfnamefont {A.~D.}\
  \bibnamefont {Bandrauk}}, \bibinfo {author} {\bibfnamefont {F.}~\bibnamefont
  {Remacle}}, \ and\ \bibinfo {author} {\bibfnamefont {H.~J.}\ \bibnamefont
  {W{\"o}rner}},\ }\bibfield  {title} {\enquote {\bibinfo {title} {Measurement
  and laser control of attosecond charge migration in ionized iodoacetylene},}\
  }\href {\doibase 10.1126/science.aab2160} {\bibfield  {journal} {\bibinfo
  {journal} {Science}\ }\textbf {\bibinfo {volume} {350}},\ \bibinfo {pages}
  {790--795} (\bibinfo {year} {2015})}\BibitemShut {NoStop}%
\bibitem [{\citenamefont {Krause}\ \emph {et~al.}(1992)\citenamefont {Krause},
  \citenamefont {Schafer},\ and\ \citenamefont {Kulander}}]{Krause_1992}%
  \BibitemOpen
  \bibfield  {author} {\bibinfo {author} {\bibfnamefont {Jeffrey~L.}\
  \bibnamefont {Krause}}, \bibinfo {author} {\bibfnamefont {Kenneth~J.}\
  \bibnamefont {Schafer}}, \ and\ \bibinfo {author} {\bibfnamefont
  {Kenneth~C.}\ \bibnamefont {Kulander}},\ }\bibfield  {title} {\enquote
  {\bibinfo {title} {High-order harmonic generation from atoms and ions in the
  high intensity regime},}\ }\href {\doibase 10.1103/PhysRevLett.68.3535}
  {\bibfield  {journal} {\bibinfo  {journal} {Phys. Rev. Lett.}\ }\textbf
  {\bibinfo {volume} {68}},\ \bibinfo {pages} {3535--3538} (\bibinfo {year}
  {1992})}\BibitemShut {NoStop}%
\bibitem [{\citenamefont {Corkum}(1993)}]{Corkum_HHG}%
  \BibitemOpen
  \bibfield  {author} {\bibinfo {author} {\bibfnamefont {P.~B.}\ \bibnamefont
  {Corkum}},\ }\bibfield  {title} {\enquote {\bibinfo {title} {Plasma
  perspective on strong field multiphoton ionization},}\ }\href {\doibase
  10.1103/PhysRevLett.71.1994} {\bibfield  {journal} {\bibinfo  {journal}
  {Phys. Rev. Lett.}\ }\textbf {\bibinfo {volume} {71}},\ \bibinfo {pages}
  {1994--1997} (\bibinfo {year} {1993})}\BibitemShut {NoStop}%
\bibitem [{\citenamefont {Lewenstein}\ \emph {et~al.}(1994)\citenamefont
  {Lewenstein}, \citenamefont {Balcou}, \citenamefont {Ivanov}, \citenamefont
  {L'Huillier},\ and\ \citenamefont {Corkum}}]{Lewenstein_HHG}%
  \BibitemOpen
  \bibfield  {author} {\bibinfo {author} {\bibfnamefont {M.}~\bibnamefont
  {Lewenstein}}, \bibinfo {author} {\bibfnamefont {Ph.}\ \bibnamefont
  {Balcou}}, \bibinfo {author} {\bibfnamefont {M.~Yu.}\ \bibnamefont {Ivanov}},
  \bibinfo {author} {\bibfnamefont {Anne}\ \bibnamefont {L'Huillier}}, \ and\
  \bibinfo {author} {\bibfnamefont {P.~B.}\ \bibnamefont {Corkum}},\ }\bibfield
   {title} {\enquote {\bibinfo {title} {Theory of high-harmonic generation by
  low-frequency laser fields},}\ }\href {\doibase 10.1103/PhysRevA.49.2117}
  {\bibfield  {journal} {\bibinfo  {journal} {Phys. Rev. A}\ }\textbf {\bibinfo
  {volume} {49}},\ \bibinfo {pages} {2117--2132} (\bibinfo {year}
  {1994})}\BibitemShut {NoStop}%
\bibitem [{\citenamefont {Hentschel}\ \emph {et~al.}(2001)\citenamefont
  {Hentschel}, \citenamefont {Kienberger}, \citenamefont {Spielmann},
  \citenamefont {Reider}, \citenamefont {Milosevic}, \citenamefont {Brabec},
  \citenamefont {Corkum}, \citenamefont {Heinzmann}, \citenamefont {Drescher},\
  and\ \citenamefont {Krausz}}]{Isolated_atto_pulse}%
  \BibitemOpen
  \bibfield  {author} {\bibinfo {author} {\bibfnamefont {M.}~\bibnamefont
  {Hentschel}}, \bibinfo {author} {\bibfnamefont {R.}~\bibnamefont
  {Kienberger}}, \bibinfo {author} {\bibfnamefont {Ch}~\bibnamefont
  {Spielmann}}, \bibinfo {author} {\bibfnamefont {G.~A.}\ \bibnamefont
  {Reider}}, \bibinfo {author} {\bibfnamefont {N.}~\bibnamefont {Milosevic}},
  \bibinfo {author} {\bibfnamefont {T.}~\bibnamefont {Brabec}}, \bibinfo
  {author} {\bibfnamefont {P.}~\bibnamefont {Corkum}}, \bibinfo {author}
  {\bibfnamefont {U.}~\bibnamefont {Heinzmann}}, \bibinfo {author}
  {\bibfnamefont {M.}~\bibnamefont {Drescher}}, \ and\ \bibinfo {author}
  {\bibfnamefont {F.}~\bibnamefont {Krausz}},\ }\bibfield  {title} {\enquote
  {\bibinfo {title} {Attosecond metrology},}\ }\href {\doibase
  10.1038/35107000} {\bibfield  {journal} {\bibinfo  {journal} {Nature}\
  }\textbf {\bibinfo {volume} {414}},\ \bibinfo {pages} {509--513} (\bibinfo
  {year} {2001})}\BibitemShut {NoStop}%
\bibitem [{\citenamefont {Popmintchev}\ \emph {et~al.}(2012)\citenamefont
  {Popmintchev}, \citenamefont {Chen}, \citenamefont {Popmintchev},
  \citenamefont {Arpin}, \citenamefont {Brown}, \citenamefont {Ali{\v
  s}auskas}, \citenamefont {Andriukaitis}, \citenamefont {Bal{\v c}iunas},
  \citenamefont {M{\"u}cke}, \citenamefont {Pugzlys}, \citenamefont {Baltu{\v
  s}ka}, \citenamefont {Shim}, \citenamefont {Schrauth}, \citenamefont {Gaeta},
  \citenamefont {Hern{\'a}ndez-Garc{\'\i}a}, \citenamefont {Plaja},
  \citenamefont {Becker}, \citenamefont {Jaron-Becker}, \citenamefont
  {Murnane},\ and\ \citenamefont {Kapteyn}}]{atto_pulse_creation2}%
  \BibitemOpen
  \bibfield  {author} {\bibinfo {author} {\bibfnamefont {Tenio}\ \bibnamefont
  {Popmintchev}}, \bibinfo {author} {\bibfnamefont {Ming-Chang}\ \bibnamefont
  {Chen}}, \bibinfo {author} {\bibfnamefont {Dimitar}\ \bibnamefont
  {Popmintchev}}, \bibinfo {author} {\bibfnamefont {Paul}\ \bibnamefont
  {Arpin}}, \bibinfo {author} {\bibfnamefont {Susannah}\ \bibnamefont {Brown}},
  \bibinfo {author} {\bibfnamefont {Skirmantas}\ \bibnamefont {Ali{\v
  s}auskas}}, \bibinfo {author} {\bibfnamefont {Giedrius}\ \bibnamefont
  {Andriukaitis}}, \bibinfo {author} {\bibfnamefont {Tadas}\ \bibnamefont
  {Bal{\v c}iunas}}, \bibinfo {author} {\bibfnamefont {Oliver~D.}\ \bibnamefont
  {M{\"u}cke}}, \bibinfo {author} {\bibfnamefont {Audrius}\ \bibnamefont
  {Pugzlys}}, \bibinfo {author} {\bibfnamefont {Andrius}\ \bibnamefont
  {Baltu{\v s}ka}}, \bibinfo {author} {\bibfnamefont {Bonggu}\ \bibnamefont
  {Shim}}, \bibinfo {author} {\bibfnamefont {Samuel~E.}\ \bibnamefont
  {Schrauth}}, \bibinfo {author} {\bibfnamefont {Alexander}\ \bibnamefont
  {Gaeta}}, \bibinfo {author} {\bibfnamefont {Carlos}\ \bibnamefont
  {Hern{\'a}ndez-Garc{\'\i}a}}, \bibinfo {author} {\bibfnamefont {Luis}\
  \bibnamefont {Plaja}}, \bibinfo {author} {\bibfnamefont {Andreas}\
  \bibnamefont {Becker}}, \bibinfo {author} {\bibfnamefont {Agnieszka}\
  \bibnamefont {Jaron-Becker}}, \bibinfo {author} {\bibfnamefont {Margaret~M.}\
  \bibnamefont {Murnane}}, \ and\ \bibinfo {author} {\bibfnamefont {Henry~C.}\
  \bibnamefont {Kapteyn}},\ }\bibfield  {title} {\enquote {\bibinfo {title}
  {Bright coherent ultrahigh harmonics in the kev x-ray regime from
  mid-infrared femtosecond lasers},}\ }\href {\doibase 10.1126/science.1218497}
  {\bibfield  {journal} {\bibinfo  {journal} {Science}\ }\textbf {\bibinfo
  {volume} {336}},\ \bibinfo {pages} {1287--1291} (\bibinfo {year}
  {2012})}\BibitemShut {NoStop}%
\bibitem [{\citenamefont {Koval}\ \emph {et~al.}(2007)\citenamefont {Koval},
  \citenamefont {Wilken}, \citenamefont {Bauer},\ and\ \citenamefont
  {Keitel}}]{NSDR}%
  \BibitemOpen
  \bibfield  {author} {\bibinfo {author} {\bibfnamefont {P.}~\bibnamefont
  {Koval}}, \bibinfo {author} {\bibfnamefont {F.}~\bibnamefont {Wilken}},
  \bibinfo {author} {\bibfnamefont {D.}~\bibnamefont {Bauer}}, \ and\ \bibinfo
  {author} {\bibfnamefont {C.~H.}\ \bibnamefont {Keitel}},\ }\bibfield  {title}
  {\enquote {\bibinfo {title} {Nonsequential double recombination in intense
  laser fields},}\ }\href {\doibase 10.1103/PhysRevLett.98.043904} {\bibfield
  {journal} {\bibinfo  {journal} {Phys. Rev. Lett.}\ }\textbf {\bibinfo
  {volume} {98}},\ \bibinfo {pages} {043904} (\bibinfo {year}
  {2007})}\BibitemShut {NoStop}%
\bibitem [{\citenamefont {Hansen}\ and\ \citenamefont {Madsen}(2016)}]{SPEAR}%
  \BibitemOpen
  \bibfield  {author} {\bibinfo {author} {\bibfnamefont {Kenneth~K.}\
  \bibnamefont {Hansen}}\ and\ \bibinfo {author} {\bibfnamefont {Lars~Bojer}\
  \bibnamefont {Madsen}},\ }\bibfield  {title} {\enquote {\bibinfo {title}
  {Same-period emission and recombination in nonsequential double-recombination
  high-order-harmonic generation},}\ }\href {\doibase
  10.1103/PhysRevA.93.053427} {\bibfield  {journal} {\bibinfo  {journal} {Phys.
  Rev. A}\ }\textbf {\bibinfo {volume} {93}},\ \bibinfo {pages} {053427}
  (\bibinfo {year} {2016})}\BibitemShut {NoStop}%
\bibitem [{\citenamefont {Zuo}\ and\ \citenamefont {Bandrauk}(1995)}]{CREI}%
  \BibitemOpen
  \bibfield  {author} {\bibinfo {author} {\bibfnamefont {T.}~\bibnamefont
  {Zuo}}\ and\ \bibinfo {author} {\bibfnamefont {A.~D.}\ \bibnamefont
  {Bandrauk}},\ }\bibfield  {title} {\enquote {\bibinfo {title}
  {Charge-resonance-enhanced ionization of diatomic molecular ions by intense
  lasers},}\ }\href {\doibase 10.1103/PhysRevA.52.R2511} {\bibfield  {journal}
  {\bibinfo  {journal} {Phys. Rev. A}\ }\textbf {\bibinfo {volume} {52}},\
  \bibinfo {pages} {R2511--R2514} (\bibinfo {year} {1995})}\BibitemShut
  {NoStop}%
\bibitem [{\citenamefont {Bandrauk}\ \emph {et~al.}(1997)\citenamefont
  {Bandrauk}, \citenamefont {Chelkowski}, \citenamefont {Yu},\ and\
  \citenamefont {Constant}}]{Bandrauk}%
  \BibitemOpen
  \bibfield  {author} {\bibinfo {author} {\bibfnamefont {A.~D.}\ \bibnamefont
  {Bandrauk}}, \bibinfo {author} {\bibfnamefont {S.}~\bibnamefont
  {Chelkowski}}, \bibinfo {author} {\bibfnamefont {H.}~\bibnamefont {Yu}}, \
  and\ \bibinfo {author} {\bibfnamefont {E.}~\bibnamefont {Constant}},\
  }\bibfield  {title} {\enquote {\bibinfo {title} {Enhanced harmonic generation
  in extended molecular systems by two-color excitation},}\ }\href {\doibase
  10.1103/PhysRevA.56.R2537} {\bibfield  {journal} {\bibinfo  {journal} {Phys.
  Rev. A}\ }\textbf {\bibinfo {volume} {56}},\ \bibinfo {pages} {R2537--R2540}
  (\bibinfo {year} {1997})}\BibitemShut {NoStop}%
\bibitem [{\citenamefont {Lein}\ \emph {et~al.}(2002)\citenamefont {Lein},
  \citenamefont {Hay}, \citenamefont {Velotta}, \citenamefont {Marangos},\ and\
  \citenamefont {Knight}}]{mol_interferens}%
  \BibitemOpen
  \bibfield  {author} {\bibinfo {author} {\bibfnamefont {M.}~\bibnamefont
  {Lein}}, \bibinfo {author} {\bibfnamefont {N.}~\bibnamefont {Hay}}, \bibinfo
  {author} {\bibfnamefont {R.}~\bibnamefont {Velotta}}, \bibinfo {author}
  {\bibfnamefont {J.~P.}\ \bibnamefont {Marangos}}, \ and\ \bibinfo {author}
  {\bibfnamefont {P.~L.}\ \bibnamefont {Knight}},\ }\bibfield  {title}
  {\enquote {\bibinfo {title} {Interference effects in high-order harmonic
  generation with molecules},}\ }\href {\doibase 10.1103/PhysRevA.66.023805}
  {\bibfield  {journal} {\bibinfo  {journal} {Phys. Rev. A}\ }\textbf {\bibinfo
  {volume} {66}},\ \bibinfo {pages} {023805} (\bibinfo {year}
  {2002})}\BibitemShut {NoStop}%
\bibitem [{\citenamefont {Torres}\ \emph {et~al.}(2007)\citenamefont {Torres},
  \citenamefont {Kajumba}, \citenamefont {Underwood}, \citenamefont {Robinson},
  \citenamefont {Baker}, \citenamefont {Tisch}, \citenamefont {de~Nalda},
  \citenamefont {Bryan}, \citenamefont {Velotta}, \citenamefont {Altucci},
  \citenamefont {Turcu},\ and\ \citenamefont {Marangos}}]{HHG_mol_probe1}%
  \BibitemOpen
  \bibfield  {author} {\bibinfo {author} {\bibfnamefont {R.}~\bibnamefont
  {Torres}}, \bibinfo {author} {\bibfnamefont {N.}~\bibnamefont {Kajumba}},
  \bibinfo {author} {\bibfnamefont {Jonathan~G.}\ \bibnamefont {Underwood}},
  \bibinfo {author} {\bibfnamefont {J.~S.}\ \bibnamefont {Robinson}}, \bibinfo
  {author} {\bibfnamefont {S.}~\bibnamefont {Baker}}, \bibinfo {author}
  {\bibfnamefont {J.~W.~G.}\ \bibnamefont {Tisch}}, \bibinfo {author}
  {\bibfnamefont {R.}~\bibnamefont {de~Nalda}}, \bibinfo {author}
  {\bibfnamefont {W.~A.}\ \bibnamefont {Bryan}}, \bibinfo {author}
  {\bibfnamefont {R.}~\bibnamefont {Velotta}}, \bibinfo {author} {\bibfnamefont
  {C.}~\bibnamefont {Altucci}}, \bibinfo {author} {\bibfnamefont {I.~C.~E.}\
  \bibnamefont {Turcu}}, \ and\ \bibinfo {author} {\bibfnamefont {J.~P.}\
  \bibnamefont {Marangos}},\ }\bibfield  {title} {\enquote {\bibinfo {title}
  {Probing orbital structure of polyatomic molecules by high-order harmonic
  generation},}\ }\href {\doibase 10.1103/PhysRevLett.98.203007} {\bibfield
  {journal} {\bibinfo  {journal} {Phys. Rev. Lett.}\ }\textbf {\bibinfo
  {volume} {98}},\ \bibinfo {pages} {203007} (\bibinfo {year}
  {2007})}\BibitemShut {NoStop}%
\bibitem [{\citenamefont {Li}\ \emph {et~al.}(2008)\citenamefont {Li},
  \citenamefont {Zhou}, \citenamefont {Lock}, \citenamefont {Patchkovskii},
  \citenamefont {Stolow}, \citenamefont {Kapteyn},\ and\ \citenamefont
  {Murnane}}]{HHG_mol_probe2}%
  \BibitemOpen
  \bibfield  {author} {\bibinfo {author} {\bibfnamefont {Wen}\ \bibnamefont
  {Li}}, \bibinfo {author} {\bibfnamefont {Xibin}\ \bibnamefont {Zhou}},
  \bibinfo {author} {\bibfnamefont {Robynne}\ \bibnamefont {Lock}}, \bibinfo
  {author} {\bibfnamefont {Serguei}\ \bibnamefont {Patchkovskii}}, \bibinfo
  {author} {\bibfnamefont {Albert}\ \bibnamefont {Stolow}}, \bibinfo {author}
  {\bibfnamefont {Henry~C.}\ \bibnamefont {Kapteyn}}, \ and\ \bibinfo {author}
  {\bibfnamefont {Margaret~M.}\ \bibnamefont {Murnane}},\ }\bibfield  {title}
  {\enquote {\bibinfo {title} {Time-resolved dynamics in n2o4 probed using high
  harmonic generation},}\ }\href {\doibase 10.1126/science.1163077} {\bibfield
  {journal} {\bibinfo  {journal} {Science}\ }\textbf {\bibinfo {volume}
  {322}},\ \bibinfo {pages} {1207--1211} (\bibinfo {year} {2008})}\BibitemShut
  {NoStop}%
\end{thebibliography}
\end{document}